# MAPEL: Achieving Global Optimality for a Non-convex Wireless Power Control Problem

Liping Qian, *Student Member, IEEE,* Ying Jun (Angela) Zhang, *Member, IEEE,* and Jianwei Huang, *Member, IEEE*
Department of Information Engineering, The Chinese University of Hong Kong
Shatin, New Territories, Hong Kong
{lpqian6, yjzhang, jwhuang}@ie.cuhk.edu.hk

**Abstract**

Achieving weighted throughput maximization (WTM) through power control has been a long standing open problem in interference-limited wireless networks. The complicated coupling between the mutual interferences of links gives rise to a non-convex optimization problem. Previous work has considered the WTM problem in the high signal to interference-and-noise ratio (SINR) regime, where the problem can be approximated and transformed into a convex optimization problem through proper change of variables. In the general SINR regime, however, the approximation and transformation approach does not work. This paper proposes an algorithm, MAPEL, which globally converges to a global optimal solution of the WTM problem in the general SINR regime. The MAPEL algorithm is designed based on three key observations of the WTM problem: (1) the objective function is monotonically increasing in SINR, (2) the objective function can be transformed into a product of exponentiated linear fraction functions, and (3) the feasible set of the equivalent transformed problem is always "normal" although not necessarily convex. The MAPLE algorithm finds the desired optimal power control solution by constructing a series of polyblocks that approximate the feasible SINR region in increasing precision. Furthermore, by tuning the approximation factor in MAPEL, we could engineer a desirable tradeoff between optimality and convergence time. MAPEL provides an important benchmark for performance evaluation of other heuristic algorithms targeting the same problem. With the help of MAPEL, we evaluate the performance of several respective algorithms through extensive simulations.

# I. Introduction

Due to the broadcast nature of wireless communications, simultaneous transmissions in the same channel interfere with each other and limit the wireless network performance. One important interference mitigation technique is transmit-power control at the physical layer. This technique has been well studied and implemented in the context of wireless cellular communications (see a recent survey in [1]). The research in this area can be divided into two main threads. The first thread is concerned with achieving fixed signal to interference-plus-noise ratio (SINR) targets with minimum transmission power (e.g., [2]-[9]). This formulation is motivated by traditional voice communications, where an SINR higher than the threshold is not useful in terms of user perceived Quality of Service (QoS). The second thread is concerned with joint SINR allocation and power control. This formulation is motivated by data communication applications, where higher SINR means higher data rate and better QoS. Such joint optimization becomes more important as data applications will be dominant in next generation wireless networks (e.g., 4G all IP-based communication systems).

The joint SINR allocation and power control problem is more difficult to solve than the fixed SINR target case. This is because we need to optimize over the entire feasible SINR region, which is typically non-convex due to complicated interference coupling between links. One important instance of this joint optimization is weighted throughput maximization (WTM), where the objective function to be maximized is $\sum_i w_i \log_2(1+\mathrm{SINR}_i)$. Here $w_i$ is user i's weight and $\log_2(1+\mathrm{SINR}_i)$ is user $i$'s achievable date rate (bps/Hz). Researchers have spent significant amount of efforts on studying this WTM problem in the past. For example, the authors in [11] considered the high SINR regime where the SINR of each link is much larger than 0dB, in which case the individual data rate can be approximated by $\log_2(\mathrm{SINR}_i)$. Under such approximation, the WTM problem can be transformed into a convex one in the form of geometric programming (GP) by proper change of variables, and thus can be solved efficiently in a centralized fashion. A different approach was considered in [8], where the authors showed that the feasible SINR region is convex in the logarithm of SINR. This also explains why the approximation and convexification in [11] is suitable under the high SINR regime. Unfortunately, the high-SINR assumption is not valid in general for practical wireless ad-hoc networks when nearby links heavily interfere with each other. As a result, standard GP often yields a solution that is far from optimum due to possible strong interferences between links nearby. Compared with GP, the work [12] does not require high-SINR assumption. In particular, the authors in [12] first transformed the WTM problem into an equivalent signomial programming (SP), which is provably NP hard to solve. Then the authors adopted a successive convex programming method, SP Condensation (SPC) algorithm to solve SP. Similar to many algorithms used to solve non-convex optimization, the SPC algorithm only guarantees local optimal solutions. An improper initialization may considerably degrade the system throughput. To date, achieving a global optimal solution of the WTM problem still is an open problem.

In this paper, we propose a MAPEL (MLFP-bAsed PowEr aLlocation) algorithm, which is the first algorithm in the literature that can achieve the global optimal solution of the WTM problem in the general SINR regime. There are three key observations that enable MAPEL to efficiently solve the non-convex optimization problem. First, the objective function of WTM is monotonically increasing in SINR. This means the optimal solution is achieved at the boundary of the feasible SINR region. Second, the objective function of WTM can be transformed into a product of exponentiated linear fractional functions, which can be further formulated into a multiplicative linear fractional programming (MLFP) problem with nice computational features. Last, the feasible set of the equivalent transformed problem, although may be not convex, is always *normal*[1]**.** This, together with monotonicity, allows us to construct a sequence of polyblocks to approximate that SINR region boundary with increasing level of accuracy. Given an arbitrary small and finite error tolerance level, MAPEL is guaranteed to find one global optimal solution of the WMT problem within finite amount of time. A flexible tradeoff between performance and convergence time can be achieved by tuning the approximation factor.

The main benefit of MAPEL is to provide a benchmark for all algorithms that are designed to tackle the WMT problem, whether it is existing or to be proposed, centralized or distributed, optimal or heuristic. In this paper, we show how such benchmark is useful in elevating the performance of two state-of-art centralized and distributed algorithms ([12], [16]) in this area.

Finally, we note that some work has been done on the problem of maximizing the minimum achievable SINR of each link in wireless networks [4], [10]. This is motivated partially by fair allocation among various users in the network. All existing algorithms for solving this problem are centralized. Interestingly, our MAPEL algorithm can be easily adapted to solve the same max-min optimization problem in a different and also centralized manner. We will briefly discuss this extension as well.

The remainder of this paper is organized as follows. System model is discussed in Section II. In Section III, we transform the throughput-maximization power control problem into a MLFP problem. Some properties of the feasible region in MLFP problem are also discussed. The MAPEL algorithm is proposed and analyzed in Section IV. A brief discussion on the extension to the max-min SINR problem is also provided. In Section V, we evaluate the performance of MAPEL through several simulations. With the benchmark established by MAPEL, we evaluate the performance of two existing algorithms in Section VI. The paper is concluded in Section VII.

---

[1] Various math preliminaries and definitions are given in Section V.

Throughout the paper, vectors are denoted in bold small letter, e.g., $\mathbf{z}$, with its $i$th component $z_i$. Matrices are denoted by bold capitalized letters, e.g., $\boldsymbol{Z}$, with $Z_{ij}$ denoting the $\{i, j\}$th component. Sets are denoted by Euler letters, e.g., $\mathcal{A}$.

## II. System Model

We consider a wireless ad hoc network with a set of $\mathcal{M} = \{1, \cdots, M\}$ *distinct* links[2]. Each link consists of a transmitter node $T_i$ and a receiver node $R_i$. The channel gain between node $T_i$ and node $R_j$ is denoted by $G_{ij}$, which is determined by various factors such as path loss, shadowing and fading effects. The complete channel matrix is denoted by $\boldsymbol{G} = [G_{ij}]$. Let $p_i$ denote the transmitting power of link $i$ (i.e., from node $T_i$), and $n_i$ denote the receiving noise on link $i$ (i.e., measured at node $R_i$). The received signal to interference-plus-noise ratio (SINR) of link $i$ is

$$\gamma_i(\mathbf{p}) = \frac{G_{ii} p_i}{\sum_{j \neq i} G_{ji} p_j + n_i}, \tag{1}$$

and the data rate calculated based on the Shannon capacity formula is $\log_2(1+\gamma_i(\mathbf{p}))$[3]. To simplify notations, we use $\boldsymbol{p} = (p_i, \forall i \in \mathcal{M})$ and $\boldsymbol{\gamma}(\boldsymbol{p}) = (\gamma_i(\boldsymbol{p}), \forall i \in \mathcal{M})$ to represent the transmission power vector and achieved SINR vector of all links.

We want to find the optimal power allocation $\boldsymbol{p}^*$ that maximizes the weighted sum throughput subject to individual data rate constraints. Mathematically we want to solve the following optimization problem

$$\begin{aligned} \underset{\boldsymbol{p}}{\text{maximize}} \quad & \sum_{i=1}^{M} w_i \log_2(1+\gamma_i(\boldsymbol{p})) \\ \text{subject to} \quad & \log_2(1+\gamma_i(\boldsymbol{p})) \geq r_{i,\min}, \forall i \in \mathcal{M}, \\ & 0 \leq p_i \leq P_i^{max}, \forall i \in \mathcal{M}. \end{aligned} \tag{P1}$$

Here $r_{i,\min} \geq 0$ is the minimum data rate requirement of link $i$ (including the special case of $r_{i,\min}=0$, i.e., no rate constraint), and $w_i > 0$ is the priority weight of link $i$. Without loss of generality, the weights $w_i$ are normalized so that $\sum_{i=1}^{M} w_i = 1$. Notice that if $r_{i,\min}$'s are too large, there may not exist a feasible solution.

---

[2] For example,this could represent a network snapshot under a particular schedule of transmissions determined by an underlying routing and MAC protocol.

[3] To better model the achievable rates in a practical system, we can re-normalize $\gamma_i$ by $\beta\gamma_i$, where $\beta \in [0,1]$ represents the system's "gap" from capacity. Such modification, however, does not change the analysis in this paper.

For a user *i*, its received SINR value needs to be at least $\gamma_{i,\min} = 2^{r_{i,\min}} - 1$ in order to satisfy its minimum rate requirement. Consider the following matrix $\boldsymbol{B}$,

$$B_{ij} = \begin{cases} 0, & i = j \\ \frac{\gamma_{i,\min} G_{ji}}{G_{ii}}, & i \neq j \end{cases}.$$

According to Theorem 2.1 in [1], if the maximum eigenvalue of $\boldsymbol{B}$ is larger than 1, then there is no feasible solution to Problem P1. Otherwise, we can find a power allocation $\hat{\boldsymbol{p}}$ as follows,

$$\hat{\boldsymbol{p}} = (\boldsymbol{I} - \boldsymbol{B})^{-1} \boldsymbol{u},$$

where $\boldsymbol{I}$ is the $M \times M$ identity matrix and $\boldsymbol{u}$ is a $M \times 1$ vector with elements

$$u_i = \frac{\gamma_{i,\min} n_i}{G_{ii}} .$$

By Theorem 2.2 in [1], Problem P1 is feasible if and only if the components of $\hat{\boldsymbol{p}}$ satisfies $0 \le \hat{p}_i \le P_i^{max}$ for all $i$,. Therefore, the procedure of checking the feasibility of Problem P1 is as follows:

---

**Procedure 1** Check the feasibility of $r_{i,\min}$'s

---

1: Transform minimum data constraints into minimum SINR constraints through $\gamma_{i,\min}(\boldsymbol{p}) = 2^{r_{i,\min}} - 1$ for all $i$.
2: Compute the maximum eigenvalue of matrix $\boldsymbol{B}$ and check if it is smaller than 1. If not, $r_{i,\min}$'s are infeasible. Otherwise, go to step 3.
3: Compute the power allocation $\hat{\boldsymbol{p}} = (\boldsymbol{I} - \boldsymbol{B})^{-1} \boldsymbol{u}$ and check if it satisfies $0 \le \hat{p}_i \le P_i^{max}$ for all $i$. If so, $r_{i,\min}$'s are feasible. Otherwise, $r_{i,\min}$'s are infeasible.

---

It has been shown that (e.g., [11], [12], [16], [17]) Problem P1 is a non-convex optimization problem in terms of the transmit power $\boldsymbol{p}$. Thus, it is difficult to find a global optimal solution efficiently even in a centralized fashion. In Section III, we will show Problem P1 can be transformed into a Multiplicative Linear Fractional Programming (MLFP) problem, which can then be solved efficiently by the MAPEL algorithm presented in Section IV.

## III. Power Control as Multiplicative Linear Fractional Programming (MLFP)

In this section, we first introduce the definition of Generalized Linear Fractional Programming, and show that Problem P1 can be formulated as a special case of the GLFP (which we refer to as MLFP). We further discuss several key properties of the new formulation that are critical for developing the MAPEL algorithm.

*Definition 1 GLFP:* [20] An optimization problem belongs to the class of Generalized Linear Fractional Programming (GLFP) if it can be represented by one of the following two formulations:

$$\begin{array}{ll} \text{maximize} & \Phi\left(\dfrac{f_1(\boldsymbol{x})}{g_1(\boldsymbol{x})},\cdots,\dfrac{f_M(\boldsymbol{x})}{g_M(\boldsymbol{x})}\right) \\ \text{variables} & \boldsymbol{x}\in\mathcal{D} \end{array} \tag{2}$$

or

$$\begin{array}{ll} \text{minimize} & \Phi\left(\dfrac{f_1(\boldsymbol{x})}{g_1(\boldsymbol{x})},\cdots,\dfrac{f_M(\boldsymbol{x})}{g_M(\boldsymbol{x})}\right) \\ \text{variables} & \boldsymbol{x}\in\mathcal{D}, \end{array} \tag{3}$$

where the domain $\mathcal{D}$ is a nonempty polytope[45] in $\mathcal{R}^N$ , functions $f_1,\cdots,f_M,g_1,\cdots,g_M:\mathcal{R}^N\rightarrow\mathcal{R}$ are linear affine on $\mathcal{R}^N$, and function $\Phi:\mathcal{R}^M\rightarrow\mathcal{R}$ is *increasing* on $\mathcal{R}_+^M$.

By the properties of the logarithm function, we can rewrite Problem P1 as follows:

$$\begin{array}{ll} \text{maximize} & \prod_{i=1}^{M}\left(\dfrac{f_i(\boldsymbol{p})}{g_i(\boldsymbol{p})}\right)^{w_i} \\ \text{variables} & \boldsymbol{p}\in\mathcal{P}, \end{array} \tag{P2}$$

with the feasible set

$$\mathcal{P}=\{\boldsymbol{p}\,|\,0\le p_i\le P_i^{max},\frac{f_i(\boldsymbol{p})}{g_i(\boldsymbol{p})}\ge 2^{r_{i,\min}},\forall i\in\mathcal{M}\}, \tag{4}$$

which is a nonempty polytope in $\mathcal{R}^M$ . Here $f_i(\boldsymbol{p})=G_{ii}p_i+\sum_{j\neq i}G_{ji}p_j+n_i$ and $g_i(\boldsymbol{p})=\sum_{j\neq i}G_{ji}p_j+n_i$ for all $i$. It is clear that the objective function of Problem P2 is a product of exponentiated linear fractional functions, and the function $\Phi(\boldsymbol{z})=\prod_{i=1}^{M}(z_i)^{w_i}$ is an increasing

4 Polytope means the generalization to any dimension of polygon in two dimensions, polyhedron in three dimensions, and polychoron in four dimensions.

5 $\mathcal{R}^N$ denotes $N$ -dim real domain and $\mathcal{R}_+^N$ denotes $N$ -dim non-negative domain.

function on $\mathcal{R}_+^M$. That is, for any two vectors $z_1$ and $z_2$ such that $z_1 \succeq z_2$ [6], we have $\Phi(z_1) \ge \Phi(z_2)$. Therefore, Problem P2 is a special case of GLFP, which we refer to as Multiplicative Linear Fractional Programming (MLFP) due to the multiplicative nature of the objective function.

We further note that $f_i(\boldsymbol{p})$ and $g_i(\boldsymbol{p})$ in Problem P2 are always strictly positive due to the existence of positive noise power $n_i$. Based on this, we can further rewrite Problem P2 as

$$\begin{aligned} &\text{maximize} \quad \Phi(z) = \prod_{i=1}^{M} (z_i)^{w_i} \\ &\text{variables} \quad z \in \mathcal{G}, \end{aligned} \qquad \text{(P3)}$$

where the feasible set

$$\mathcal{G} = \{z \mid 0 \le z_i \le \frac{f_i(\boldsymbol{p})}{g_i(\boldsymbol{p})}, \forall i \in \mathcal{M}, \boldsymbol{p} \in \mathcal{P}\}. \qquad (6)$$

Since $\Phi(z)$ is an increasing function in $z$, the optimal solution to Problem P3, denoted by $z^*$, must occur at places where $z_i = \frac{f_i(\boldsymbol{p})}{g_i(\boldsymbol{p})}$ for all $i$. If we can find a power allocation $\mathbf{p}^*$ corresponding to the optimal solution $z^*$ such that $z_i^* = \frac{f_i(\mathbf{p}^*)}{g_i(\mathbf{p}^*)}$ for all $i$, then such $\mathbf{p}^*$ is clearly the optimal solution to Problem P2. Finding such $\mathbf{p}^*$ requires solving $M$ linear equations $z_i^* g_i(\mathbf{p}^*) - f_i(\mathbf{p}^*) = 0$ with $M$ variables $p_1^*, \cdots, p_M^*$. As the coefficients of $f_i(\mathbf{p}^*)$ and $g_i(\mathbf{p}^*)$ consist of random channel gains $G_{ij}$'s, we can show with probability 1 that the $M$ equations are linearly independent, implying there is a unique solution $\mathbf{p}^*$. Hence Problems P1, P2 and P3 are all equivalent with each other. *We will focus on how to solve Problem P3 efficiently in the rest of the paper.*

Before attempting to solve Problem P3, it is critical to understand several important properties of the feasible set $\mathcal{G}$ in (5). The following definition will be useful in later discussions.

*Definition 2 (Box):* Given any vector $\boldsymbol{v} \in \mathcal{R}_+^M$, the hyper rectangle $[0, \boldsymbol{v}] = \{\boldsymbol{x} \mid 0 \preceq \boldsymbol{x} \preceq \boldsymbol{v}\}$ is a box with vertex $\boldsymbol{v}$ [7].

According to this definition, the feasible set $\mathcal{G}$ can be characterized as a union of infinite number of boxes with vertices of all boxes belonging to the set $\{\boldsymbol{c} \mid c_i = \frac{f_i(\boldsymbol{p})}{g_i(\boldsymbol{p})}, \forall i \in \mathcal{M}, \boldsymbol{p} \in \mathcal{P}\}$. Each element in this set is determined by a power vector $\boldsymbol{p}$ that is feasible in Problem P1 (and Problem

[6] In this paper, $\mathbf{a} \prec \mathbf{b}$ means $\mathbf{a}$ is component-wise smaller than $\mathbf{b}$, $\mathbf{a} \succ \mathbf{b}$ means $\mathbf{a}$ is component-wise larger than $\mathbf{b}$, $\mathbf{a} \preceq \mathbf{b}$ means $\mathbf{a}$ is component-wise smaller than or equal to $\mathbf{b}$, and $\mathbf{a} \succeq \mathbf{b}$ means $\mathbf{a}$ is component-wise larger than or equal to $\mathbf{b}$.

[7] In this paper, $0$ is a $1 \times M$ vector with every element being 0, and $1$ is a $1 \times M$ vector with every element being 1

P2).

*Definition 3 (Normal):* An infinite set $\mathcal{F} \subset \mathcal{R}_{+}^{M}$ is normal if for any element $\boldsymbol{v} \in \mathcal{F}$, the set $[0, \boldsymbol{v}] \subset \mathcal{F}$.

***Proposition*** 1: The intersection and union of a family of normal sets are normal sets.

*Remark 1:* Since the feasible set $\mathcal{G}$ of Problem P3 is the union of infinite number of boxes, it is a normal set.

Fig. 1 illustrates one possible example of the shape of $\mathcal{G}$ in a 2-link network. Note that $\mathcal{G}$ is in general a non-convex set. However, this paper shows that convexity of the feasible set is not important in obtaining the global optimal solution. It is the monotonicity of the objective function in the reformulated problem P3 that facilitates efficient calculation of the global optimal solution.

Before leaving this section, note that $\frac{f_i(\boldsymbol{p})}{g_i(\boldsymbol{p})}$ is lower bounded by $2^{r_{i,\min}}$ for $\boldsymbol{p} \in \mathcal{P}$. Consequently, the optimal solution $\boldsymbol{z}^*$ to P3, which occurs only at places where $z_i = \frac{f_i(p)}{g_i(p)}$ for all $i$, is also lower bounded by $2^{r_{i,\min}}$. In other words, the optimal solution $\boldsymbol{z}^*$ must reside in the set $\mathcal{G} \cap \Theta$, where $\Theta = \{\boldsymbol{z} \mid z_i \succeq 2^{r_{i,\min}} \ \forall i \in \mathcal{M}\}$.

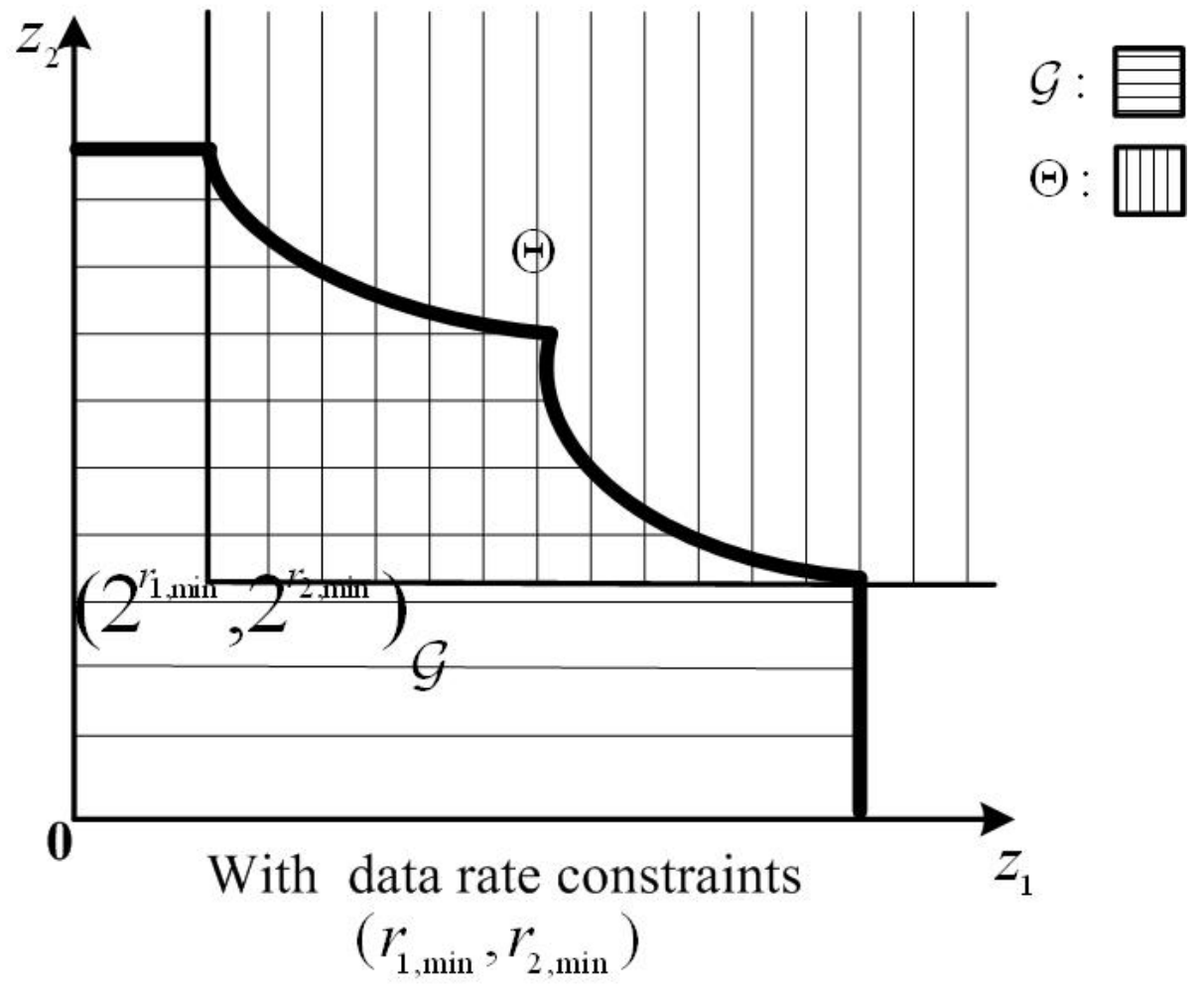

Fig. 1. Shapes of $\mathcal{G}$ and $\mathcal{L}$ for a two-link network

## IV. THE MAPEL ALGORITHM

In this section, we propose a novel algorithm, MAPEL, to solve Problem P3 based on the special characteristics of MLFP. Some mathematical preliminaries will be introduced first before we present the algorithm.

*A. Related Mathematical Preliminaries*

*Definition 4 (Polyblock):* Given any finite set $\mathcal{T} \subset \mathcal{R}_+^M$ with elements $\boldsymbol{v}$, the union of all the boxes $[0, \boldsymbol{v}_i]$ is a polyblock with vertex set $\mathcal{T}$.

*Definition 5 (Proper):* An element $\boldsymbol{v} \in \mathcal{T}$ is proper if there does not exist $\tilde{\boldsymbol{v}} \in \mathcal{T}$ such that $\boldsymbol{v} \neq \tilde{\boldsymbol{v}}$ and $\tilde{\boldsymbol{v}} \succeq \boldsymbol{v}$. In other words, a proper element is Pareto optimal. If every element $\boldsymbol{v} \in \mathcal{T}$ is proper, then the set $\mathcal{T}$ is a proper set.

***Proposition 2:*** If $\Phi(\boldsymbol{v})$: $\mathcal{R}_+^N \to \mathcal{R}_+$ is an increasing function of $\boldsymbol{v}$, then the maximum of $\Phi(\boldsymbol{v})$ over a polyblock occurs at one proper vertex of this polyblock.

*Proof:* Let $\boldsymbol{v}^*$ be a global optimal solution of $\Phi(\boldsymbol{v})$ over a polyblock $\mathcal{S}$. If $\boldsymbol{v}^*$ is not a proper vertex of $\mathcal{S}$, then $\boldsymbol{v}^* \preceq \tilde{\boldsymbol{v}}$ for some proper vertex $\tilde{\boldsymbol{v}} \neq \boldsymbol{v}^*$. Since $\Phi(\mathbf{v}^*) \leq \Phi(\tilde{\mathbf{v}})$ due to the increasing property of $\Phi(\boldsymbol{v})$, it follows that $\tilde{\boldsymbol{v}}$ is also a global optimal solution of $\Phi(\boldsymbol{v})$, which is a contradiction to $\boldsymbol{v}^*$ being a global optimal solution. Hence, Proposition 2 follows immediately. ■

*Definition 6 (Projection):* Given any nonempty normal set $\mathcal{F} \subset \mathcal{R}_+^M$ and any $\boldsymbol{v} \in \mathcal{R}_+^M \setminus \{0\}$[8], $\pi^{\mathcal{F}}(\boldsymbol{v})$ is a projection of $\boldsymbol{v}$ on $\mathcal{F}$ if $\pi^{\mathcal{F}}(\boldsymbol{v}) = \lambda \boldsymbol{v}$ with $\lambda = \max\{\alpha \mid \alpha \boldsymbol{v} \in \mathcal{F}\}$. In other words, $\pi^{\mathcal{F}}(\boldsymbol{v})$ is the unique point where the halfline from $0$ through $\boldsymbol{v}$ meets the upper boundary of $\mathcal{F}$.

*Definition 7 (Upper boundary):* A point $\mathbf{y} \in \mathcal{R}_+^M$ is an upper boundary point of a bounded normal set $\mathcal{F}$ if $\mathbf{y} \in \mathcal{F}$ while $K_{\mathbf{y}} \subset \mathcal{R}_+^M \setminus \mathcal{F}$[9]. The set of upper boundary points of $\mathcal{F}$ is the upper boundary of $\mathcal{F}$.

We illustrate the above concepts in Fig. 2(a). In Fig. 2(a), the rectangles $\boldsymbol{a}0\boldsymbol{c}\boldsymbol{v}_1$[10] and $\boldsymbol{b}0\boldsymbol{d}\boldsymbol{v}_2$ represent boxes $[0, \boldsymbol{v}_1]$ and $[0, \boldsymbol{v}_2]$, respectively. $\boldsymbol{v}_1$ and $\boldsymbol{v}_2$ are the respective vertices of these two boxes. The area consisting of rectangles $\boldsymbol{a}0\boldsymbol{c}\boldsymbol{v}_1$ and $\boldsymbol{b}0\boldsymbol{d}\boldsymbol{v}_2$ represents polyblock $\mathcal{S} = [0, \boldsymbol{v}_1] \cup [0, \boldsymbol{v}_2]$ with proper vertex set $\mathcal{T} = \{\boldsymbol{v}_1, \boldsymbol{v}_2\}$. If we choose any point $\boldsymbol{v}_3 \in \mathcal{S}$, it is clear that the rectangle $\boldsymbol{e}0\boldsymbol{f}\boldsymbol{v}_3$ belongs to polyblock $\mathcal{S}$, i.e., $[0, \boldsymbol{v}_3] \subset \mathcal{S}$. Hence, polyblock $\mathcal{S}$ is normal. Being the only intersection of the halfline from $0$ through $\boldsymbol{v}_4$ and the upper boundary of $\mathcal{S}$, $\pi^{\mathcal{S}}(\boldsymbol{v}_4)$ is a projection of $\boldsymbol{v}_4$ on $\mathcal{S}$. Moreover, if $\Phi(\boldsymbol{v})$ is an increasing function on $\mathcal{S}$, then $\Phi(\boldsymbol{v}) \le \max\{\Phi(\boldsymbol{v}_1), \Phi(\boldsymbol{v}_2)\}$ for all $\boldsymbol{v} \in \mathcal{S}$. In other words, the maximum of the increasing function $\Phi(\boldsymbol{v})$ occurs only at either $\boldsymbol{v}_1$ or $\boldsymbol{v}_2$, a proper vertex of $\Phi(\boldsymbol{v})$.

[8] In this paper, $A \setminus B$ denotes the set $\{\boldsymbol{x} \mid \boldsymbol{x} \in \mathcal{A} \text{ and } \boldsymbol{x} \notin \mathcal{B}\}$.

[9] $K_{\mathbf{y}} = \left\{\mathbf{y}' \in \mathcal{R}_+^M \middle| \mathbf{y}' \succeq \mathbf{y} \text{ and } \mathbf{y}' \neq \mathbf{y}\right\}$

[10] The rectangle is denoted using four letters in its four vertices.

Now let's use the above concepts to illustrate how we can construct a series of polyblocks that approximate a set $\mathcal{F}$ with increasing level of accuracy.

***Proposition 3:*** Let $\mathcal{S} \subset \mathcal{R}_+^M$ be a polyblock with proper vertex set $\mathcal{T}$. Also let $\mathcal{F}$ be a nonempty normal closed set that is contained in $\mathcal{S}$, i.e., $\mathcal{F} \subset \mathcal{S} \subset \mathcal{R}_+^M$. For a given vertex $\boldsymbol{v}_i \in \mathcal{T}$, let $\mathcal{T}'$ be the set obtained from $\mathcal{T}$ by replacing the vertex $\boldsymbol{v}_i$ with $M$ new vertices, $(\boldsymbol{v}_{i1}, \cdots, \boldsymbol{v}_{iM})$. Here the new vertex $\boldsymbol{v}_{ij} = \boldsymbol{v}_i - (v_{i,j} - \pi_j^{\mathcal{F}}(\boldsymbol{v}_i))\boldsymbol{e}_j$, where $\boldsymbol{e}_j$ is the $j$th unit vector of $\mathcal{R}_+^M$ [11], $v_{i,j}$ is the $j$th element of the old vertex $\boldsymbol{v}_i$, and $\pi_j^{\mathcal{F}}(\boldsymbol{v}_i)$ is the $j$th element of the projection $\pi^{\mathcal{F}}(\boldsymbol{v}_i)$. Note that some of the new vertices $(\boldsymbol{v}_{i1}, \cdots, \boldsymbol{v}_{iM})$ might not be proper. If we further remove all improper elements from set $\mathcal{T}'$ and obtain a new set $\mathcal{T}^*$, then the polyblock $\mathcal{S}^*$ with vertex set $\mathcal{T}^*$ satisfies $\mathcal{F} \subset \mathcal{S}^* \subset \mathcal{S}$. In this way, we have constructed a smaller polyblock $\mathcal{S}^*$ that still contains $\mathcal{F}$.

The detailed proof of Proposition 2 is omitted due to space limitation, and interested readers are referred to the Proposition 3 in [20].

We use Fig. 2(b) to illustrate the above procedure. As shown in Fig. 2(b), given $\mathcal{F}$ and $\mathcal{S}$ where $\mathcal{F} \subset \mathcal{S} \subset \mathcal{R}_+^2$, we can obtain a polyblock $\mathcal{S}^*$ with proper vertex set $\mathcal{T}^* = \{\boldsymbol{v}_{11}, \boldsymbol{v}_2\}$ satisfying $\mathcal{F} \subset \mathcal{S}^* \subset \mathcal{S}$. $\mathcal{T}^* = \{\boldsymbol{v}_{11}, \boldsymbol{v}_2\}$ is obtained by replacing $\boldsymbol{v}_1$ in $\mathcal{T} = \{\boldsymbol{v}_1, \boldsymbol{v}_2\}$ with $\boldsymbol{v}_{1i} = \boldsymbol{v}_1 - (v_{1,i} - \pi_i^{\mathcal{T}}(\boldsymbol{v}_1))\boldsymbol{e}_i$, $i = 1, 2$, and then deleting the improper element $\boldsymbol{v}_{12}$ from $\mathcal{T}' = \{\boldsymbol{v}_{11}, \boldsymbol{v}_{12}, \boldsymbol{v}_2\}$.

[11] In this paper, the $j$ th unit vector of $\mathcal{R}_+^M$, $\boldsymbol{e}_j$, denotes the vector whose every element is equal to zero except the $j$ th element being 1.

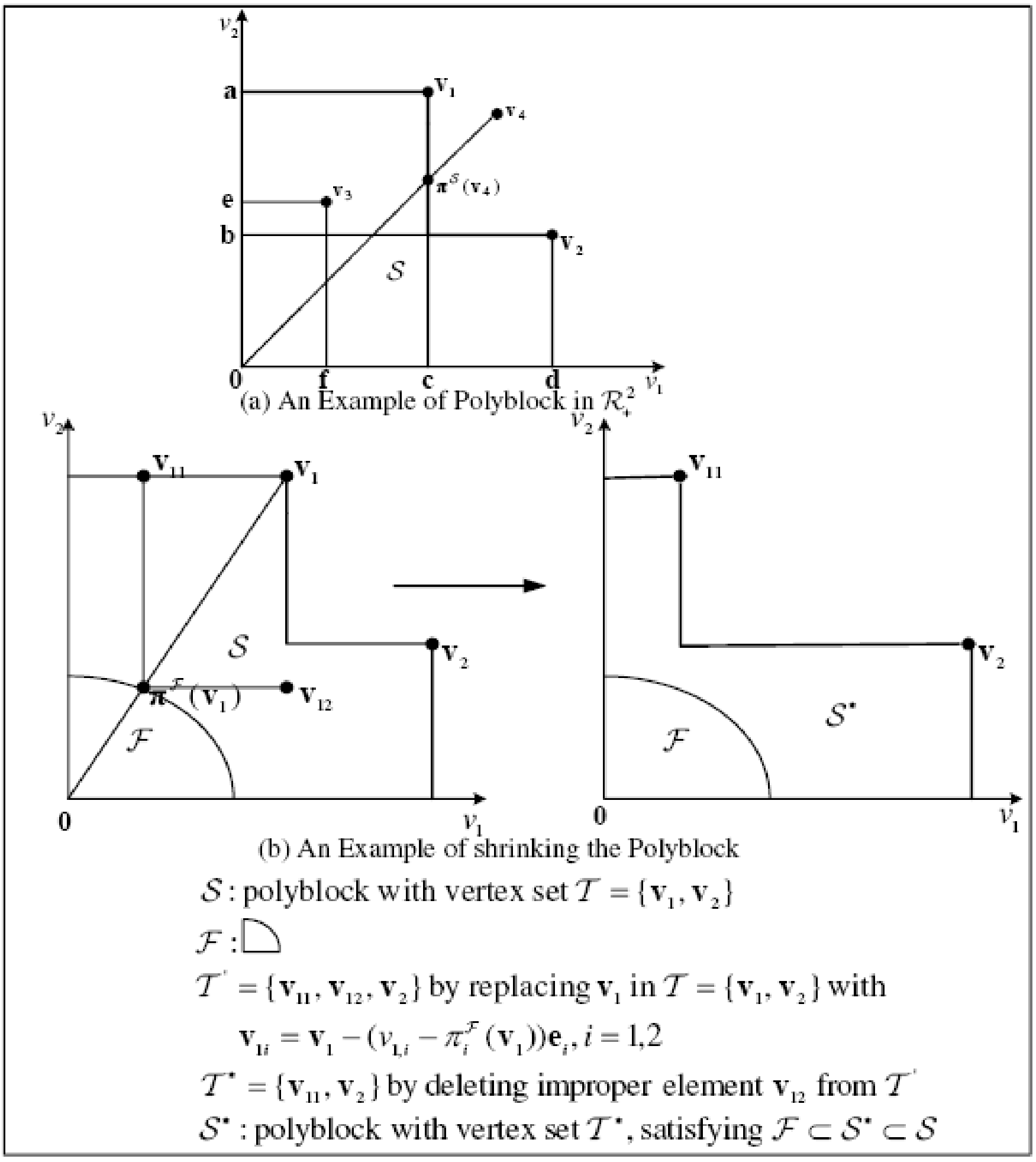

Fig. 2. Illustration about related mathematical preliminaries for MAPEL algorithm

## *B. The MAPEL Algorithm*

The MAPEL algorithm works as follows. We first construct a polyblock $\mathcal{S}_1$ that contains the feasible set of Problem P3, $\mathcal{G}$. Let $\mathcal{T}_1$ denote the proper vertex set of $\mathcal{S}_1$. By Proposition 2, the maximum of the objective function of Problem P3 (i.e., $\Phi(z) = \prod_{i=1}^{M} (z_i)^{w_i}$) over set $\mathcal{S}_1$ occurs at some proper vertex $z_1$ of $\mathcal{S}_1$, i.e., $z_1 \in \mathcal{T}_1$. If $z_1$ happens to reside in $\mathcal{G}$ as well, then it solves Problem P3 and $z^* = z_1$. Otherwise, based on Proposition 3 we can construct a smaller polyblock $\mathcal{S}_2 \subset \mathcal{S}_1$ that still contains $\mathcal{G}$ but excludes $z_1$. This is achieved by constructing the vertex set $\mathcal{T}_2$ by first replacing $z_1$ in $\mathcal{T}_1$ with $M$ new vertices $(z_{11}, \cdots, z_{1M})$, where $z_{1j} = z_1 - (z_{1,j} - \pi_j^{\mathcal{T}}(z_1))\boldsymbol{e}_j$, and then removing improper vertices. We can repeat this procedure until an optimal solution is found. This leads to a sequence of polyblocks containing $\mathcal{G}$ : $\mathcal{S}_1 \supset \mathcal{S}_2 \supset \cdots \supset \mathcal{G}$ . Obviously, $\Phi(z_1) \geq \Phi(z_2) \geq \cdots \geq \Phi(z^*)$, where $\Phi(z_i)$ is the optimal vertex that maximizes $\Phi(z)$ over set $\mathcal{S}_i$. The algorithm terminates at the $k$th iteration if $z_k \in \mathcal{G}$. For practical implementation, we say $z_k \in \mathcal{G}$ when $\max\{(z_{k,i} - \pi_i^{\mathcal{G}}(z_k))/z_{k,i}\} \leq \delta$, where $\delta > 0$ is a small positive number representing the error tolerance level.

We can further expedite the above process by selecting $z_k$ from a smaller set $\mathcal{T}_k \cap \Theta$, where $\Theta = \{z \mid z_i \succeq 2^{r_{i,\min}} \ \forall i \in \mathcal{M}\}$. This will not affect the optimality of the algorithm since the optimal solution $z^*$ is lower bounded by $2^{r_{\min}}$.

A critical step in constructing new polyblocks and checking the termination criterion is calculating the projection $\pi^{\mathcal{G}}(z_k)$. This is, however, by no means trivial, since the upper boundary of $\mathcal{G}$ is not explicitly known. In particular, $\pi^{\mathcal{G}}(z_k) = \lambda_k z_k$ is obtained by solving the following max-min problem for $\lambda_k$:

$$\begin{aligned}\lambda_k &= \max\{\alpha \mid \alpha z_k \in \mathcal{G}\} = \max\{\alpha \mid \alpha \leq \min_{1\leq i\leq M} \frac{f_i(\boldsymbol{p})}{z_{k,i} g_i(\boldsymbol{p})}, \boldsymbol{p} \in \mathcal{P}\} \\ &= \max_{p\in\mathcal{P}} \min_{1\leq i\leq M} \frac{f_i(\boldsymbol{p})}{z_{k,i} g_i(\boldsymbol{p})}.\end{aligned} \tag{6}$$

This is again a generalized linear fractional programming problem by Definition 1. We solve this problem using the Dinkelbach-type algorithm in [21] with slight modifications. The details are shown in Algorithm 1[12].

---

**Algorithm 1** Max-Min Projection Algorithm (for finding $\pi^{\mathcal{G}}(z_k)$)

---

1: **Initialization**: Choose $\boldsymbol{p}^{(0)} \in [0, \boldsymbol{P}^{max}]$ and let $j = 0$.
2: **repeat**
3: Given $\boldsymbol{p}^{(j)}$, solve $\lambda_k^{(j)} = \min_{1\leq i\leq M} \frac{f_i(\boldsymbol{p}^{(j)})}{z_{k,i} g_i(\boldsymbol{p}^{(j)})}$.
4: Given $\lambda_k^{(j)}$, solve $\boldsymbol{p}^{(j+1)} = \arg\max_{p\in\mathcal{P}} \min_{1\leq i\leq M} (f_i(\boldsymbol{p}) - \lambda_k^{(j)} z_{k,i} g_i(\boldsymbol{p}))$.
5: $j = j+1$.
6: **until** $\max_{p\in\mathcal{P}} \min_i (f_i(\boldsymbol{p}) - \lambda_k^{(j-1)} z_{k,i} g_i(\boldsymbol{p})) \leq 0$.
7: The projection is $\pi^G(z_k) = \lambda_k^{(j-1)} z_k$.

---

*Definition 8 (Q-super linear convergence):* [21] A sequence $\{s_j, j = 1, 2, \cdots\} \in \mathcal{R}$ with limit $s_\infty$ converges Q-super (quotient super) linearly if

[12] In fact, each Step 4 of Algorithm 1 is a linear programming in the convex power domain.

$$\lim_{j\to\infty}\left|\frac{s_{j+1}-s_\infty}{s_j-s_\infty}\right|=0. \tag{7}$$

***Theorem 1:*** Since $f_i(\boldsymbol{p})$ and $z_{k,i}g_i(\boldsymbol{p})$ are linear affine functions on $\boldsymbol{p}$ for all $i$ and there is a unique optimal solution to (6), the sequence $\{\lambda_k^{(j)}, j=1,2,\cdots\}$ converges Q-super linearly to the optimal solution.

*Proof*: Immediate from Theorem 8.7 in [21].

Having introduced the basic operations, we now formally present the MAPEL algorithm as follows.

---

**Algorithm 2** The MAPEL Algorithm

---

1: **Initialization**: Check the feasibility of minimum data rate requirements $r_{i,\min}$'s based on Procedure 1. If $r_{i,\min}$'s are infeasible, terminate the algorithm. Otherwise, choose the approximation factor $\delta>0$, and let $k=1$.

2: **repeat**

3: If $k=1$, construct the initial polyblock $\mathcal{S}_1$ with vertex set $\mathcal{T}_1=\{\boldsymbol{b}\}$, where the $i$ th element of vector $\boldsymbol{b}$ is

$$b_i=\max_{p\in\mathcal{P}}\frac{f_i(\boldsymbol{p})}{g_i(\boldsymbol{p})}=1+\frac{G_{ii}P_i^{max}}{n_i},\forall i\in\mathcal{M}. \tag{8}$$

It is clear that polyblock $\mathcal{S}_1$ is a box $[0,\boldsymbol{b}]$ containing $\mathcal{G}$. If $k>1$, construct a smaller polyblock $\mathcal{G}_k$ with vertex set $\mathcal{T}_k$ by replacing $\boldsymbol{z}_{k-1}$ in $\mathcal{T}_{k-1}$ with $M$ new vertices $(\boldsymbol{z}_{k-11},\cdots,\boldsymbol{z}_{k-1M})$, where $\boldsymbol{z}_{k-1j}=\boldsymbol{z}_{k-1}-(z_{k-1,j}-\pi_j^{\mathcal{G}}(\boldsymbol{z}_{k-1}))\boldsymbol{e}_j$, and removing improper vertices.

4: Find $\boldsymbol{z}_k$ that maximizes the objective function of Problem P3 over set $\mathcal{T}_k\cap\Theta$, i.e.,

$$\boldsymbol{z}_k=\operatorname{argmax}\{\Phi(\boldsymbol{z})\,|\,\boldsymbol{z}\in\mathcal{T}_k\cap\Theta\}. \tag{9}$$

5: Find $\pi^{\mathcal{G}}(\boldsymbol{z}_k)$ based on Algorithm 1.

6: $k=k+1$.

7: **until** $\max_i\{(z_{k-1,i}-\pi_i^{\mathcal{G}}(\boldsymbol{z}_{k-1}))/z_{k-1,i}\}\le\delta.$

8: Compute the optimal power allocation $\boldsymbol{p}^*$ (i.e., optimal solution to Problem P1) by solving $\pi_i^{\mathcal{G}}(\boldsymbol{z}_{k-1})=\frac{f_i(\boldsymbol{p})}{g_i(\boldsymbol{p})}$ for all $i$

---

*C. Global Convergence*

***Theorem 2:*** The MAPEL algorithm globally converges to a global optimal solution of Problem P3.

*Proof:* The MAPEL algorithm generates a sequence $\{\boldsymbol{z}_k\}$ for $k = 1, 2, \cdots$. Each component is calculated as (9) for each newly constructed polyblock. We can find a subsequence $\{\boldsymbol{z}_{k_n}\}$ within the sequence $\{\boldsymbol{z}_k\}$ such that

$$\boldsymbol{z}_{k_1} = \boldsymbol{z}_1 - (z_{1,i_1} - \pi_{i_1}^{\mathcal{G}}(\boldsymbol{z}_1))\boldsymbol{e}_{i_1}, \cdots, \boldsymbol{z}_{k_{n+1}} = \boldsymbol{z}_{k_n} - (z_{k_n,i_n} - \pi_{i_n}^{\mathcal{G}}(\boldsymbol{z}_{k_n}))\boldsymbol{e}_{i_n}, \tag{10}$$

where $1 < k_1 < k_2 < \cdots < k_n < \cdots$. $z_{k_n,i_n}$ denotes the $i_n$th element of vector $\boldsymbol{z}_{k_n}$, where $i_n$ is the only position in which $\boldsymbol{z}_{k_{n+1}}$ differs from $\boldsymbol{z}_{k_n}$. This subsequence can be thought as the "off-springs" of vertex $\boldsymbol{z}_1$ through a series of projections, and they are not necessarily adjacent since there might be projections of other vertices in between. It can be shown that there is at least one such subsequence that has infinite length. With a slight abuse of notation, let $\{\boldsymbol{z}_{k_n}, \forall n \geq 1\}$ denote such one subsequence. Since $\pi^{\mathcal{G}}(\boldsymbol{z}_{k_n}) \preceq \boldsymbol{z}_{k_n}$, (10) implies that $\boldsymbol{z}_1 \succeq \boldsymbol{z}_{k_1} \succeq \cdots \succeq \boldsymbol{z}_{k_n} \succeq \cdots \succeq 2^{r_{\min}}$. Hence, $\lim_{n\to\infty} \left\| \boldsymbol{z}_{k_n} - \boldsymbol{z}_{k_{n+1}} \right\| \to 0$. From (10) we know that $\boldsymbol{z}_{k_n}$ and $\boldsymbol{z}_{k_{n+1}}$ only differ in the $i_n$'s position, thus

$$\left\| \boldsymbol{z}_{k_n} - \boldsymbol{z}_{k_{n+1}} \right\| = z_{k_n,i_n} - z_{k_{n+1},i_n} = z_{k_n,i_n} - \pi_{i_n}^{\mathcal{G}}(\boldsymbol{z}_{k_n}) \to 0 \text{ when } n \to \infty. \tag{11}$$

Since $\pi^{\mathcal{G}}(\boldsymbol{z}_{k_n}) = \lambda_{k_n} \boldsymbol{z}_{k_n}$ and $\boldsymbol{z}_{k_n} \succeq 2^{r_{\min}}$, (11) implies that $\lim_{n\to\infty} \lambda_{k_n} = 1$. That is,

$$\lim_{n\to\infty} \boldsymbol{z}_{k_n} \to \pi^{\mathcal{G}}(\boldsymbol{z}_{k_n}). \tag{12}$$

Eqn. (12) implies that the subsequence $\{\boldsymbol{z}_{k_n}\}$ converges to the boundary of the feasible region $\mathcal{G}$. Since it is a maximizer over the set $\mathcal{S}_{k_n}$, it is thus also the global optimum of Problem P3. Note that the MAPEL algorithm terminates once the optimal solution to Problem P3 is found. Therefore, the convergence of the subsequence $\{\boldsymbol{z}_{k_n}\}$ guarantees the convergence of the algorithm to the global optimal solution. ■

*D. Trade-off between Performance and Convergence Time*

The convergence time of MAPEL is infinite if the approximation factor $\delta = 0$. However, it can be easily shown that MAPEL always terminates with finite steps when $\delta > 0$ [20]. Next, we analyze the influence of the approximation factor $\delta$ on the performance.

*Definition 9($\varepsilon-$optimal solution):* Given an $\varepsilon \geq 0$, we say that a vector $\boldsymbol{y} \in \mathcal{G}$ is an $\varepsilon-$optimal solution of Problem P3 if $\Phi(\boldsymbol{z}^*) \leq (1+\varepsilon)\Phi(\boldsymbol{y})$.

***Theorem 3:*** The solution obtained by MAPEL (if the algorithm converges) is an $\varepsilon-$optimal solution with $\varepsilon \le \frac{\delta}{1-\delta}$.

*Proof*: MAPEL terminates when $\max_i \frac{z_{k,i}-\pi_i^{\mathcal{G}}(z_k)}{z_{k,i}} \le \delta$. Consequently, together with $\sum_{i=1}^{M} w_i = 1$,

$$\Phi(\boldsymbol{z}_k)(1-\delta) \le \Phi(\pi^{\mathcal{G}}(\boldsymbol{z}_k)) \le \Phi(\boldsymbol{z}^*) \le \Phi(\boldsymbol{z}_k)$$

leading to

$$\frac{\Phi(\boldsymbol{z}_k) - \Phi(\pi^{\mathcal{G}}(\boldsymbol{z}_k))}{\Phi(\boldsymbol{z}_k)} \le \delta.$$

Note that $\Phi(\boldsymbol{z}^*) \le \Phi(\boldsymbol{z}_k)$ implies

$$\frac{\Phi(\boldsymbol{z}^*) - \Phi(\pi^{\mathcal{G}}(\boldsymbol{z}_k))}{\Phi(\boldsymbol{z}_k)} \le \delta.$$

Consequently,

$$\frac{\Phi(\boldsymbol{z}^*) - \Phi(\pi^{\mathcal{G}}(\boldsymbol{z}_k))}{\Phi(\pi^{\mathcal{G}}(\boldsymbol{z}_k))} \le \frac{\Phi(\boldsymbol{z}_k)}{\Phi(\pi^{\mathcal{G}}(\boldsymbol{z}_k))}\delta \le \frac{\delta}{1-\delta},$$

which leads to the following inequality that proves Theorem 3:

$$\Phi(z^*) \le \Phi(\pi^{\mathcal{G}}(z_k))\left(1+\frac{\delta}{1-\delta}\right). \qquad \blacksquare$$

*Remark 3:* We note that $\frac{\delta}{1-\delta} \approx \delta$ when $\delta \ll 1$. Furthermore, $\frac{\delta}{1-\delta}$ is generally a conservative estimate of $\varepsilon$. In practice, we often obtain a error that is much smaller than $\delta$.

An advantage of the MAPEL algorithm is that we can trade off performance for convergence time by tuning $\delta$. The smaller $\delta$, the longer the algorithm runs and the more accurate the optimal solution is.

### *E. Extension to Max-min SINR Power Control*

As discussed in the Introduction, some previous work on power control aimed at maximizing the minimum SINR of all links. Mathematically, they tried to solve the following problem

$$\max_{p\in\mathcal{P}} \min_i \gamma_i(\boldsymbol{p}) = \max_{p\in P} \min_i \frac{G_{ii}p_i}{\sum_{j\neq i} G_{ji}p_j + n_i} \tag{13}$$

Obviously, this is a generalized linear fractional programming defined in (2). In fact, this formulation is similar to the one in described (6). Hence, the Dinkelbach-type algorithm (Algorithm 1) that is adopted to solve (6) can be easily extended to solve the max-min SINR problem in (13).

## V. PERFORMANCE EVALUATION OF MAPEL

We illustrate the effectiveness of the MAPEL algorithm through several numerical examples.

**Example 1** (Performance and convergence time tradeoff through the approximation factor $\delta$): We consider a four-link network where the links are randomly placed in a 10m-by-10m area. The resultant channel gain matrix is

$$G_1 = \begin{bmatrix} 0.4310 & 0.0002 & 0.2605 & 0.0039 \\ 0.0002 & 0.3018 & 0.0008 & 0.0054 \\ 0.0129 & 0.0005 & 0.4266 & 0.1007 \\ 0.0011 & 0.0031 & 0.0099 & 0.0634 \end{bmatrix}. \quad (14)$$

Assume that $\boldsymbol{P}^{max}$=[0.7 0.8 0.9 1.0]mW, $n_i = 0.1\mu$W for all link $i$, and the priority weights $\boldsymbol{w} = [\frac{1}{6}\,\frac{1}{6}\,\frac{1}{3}\,\frac{1}{3}]$. Also we do not consider minimum data rate constraints in this example.

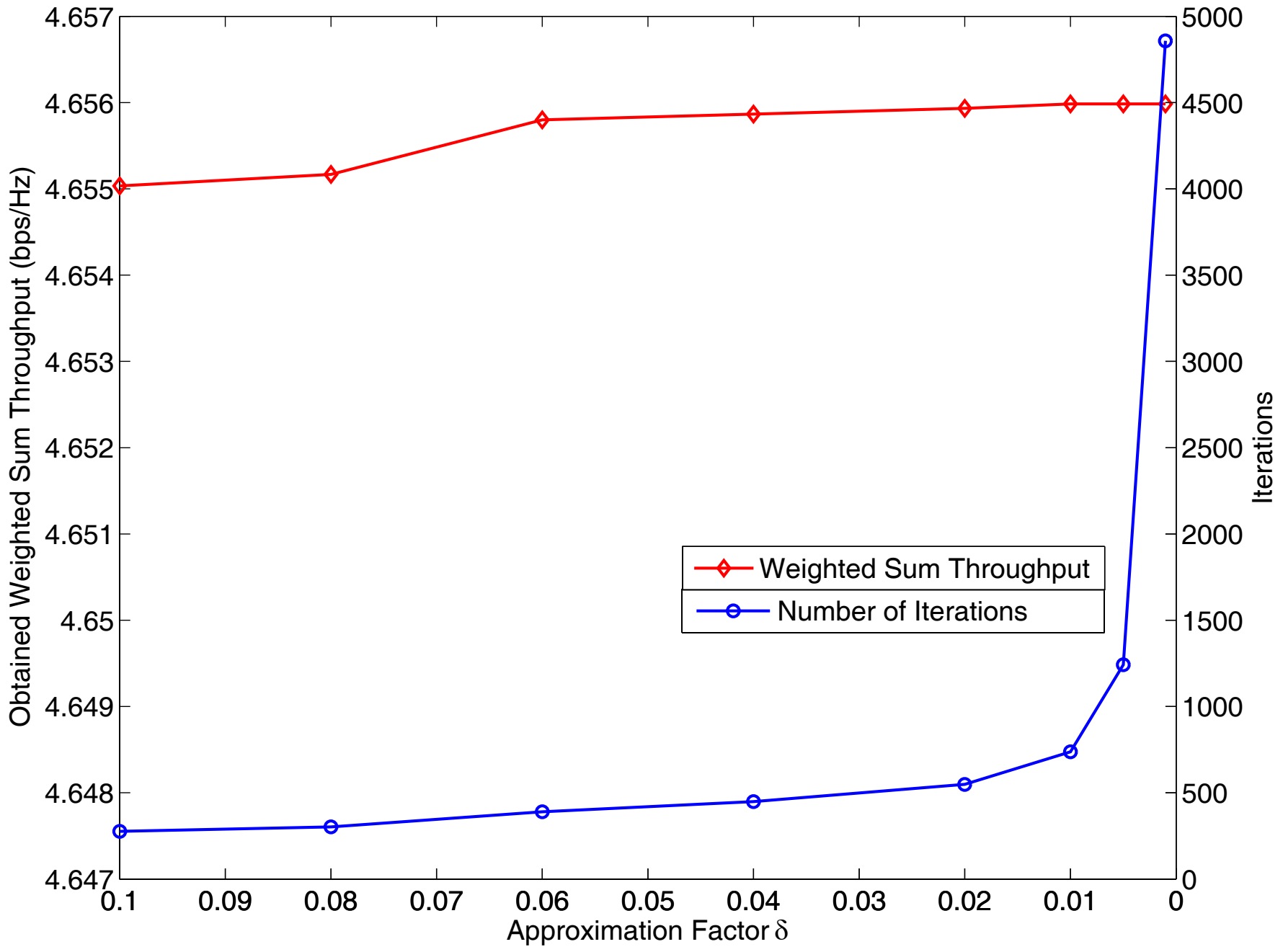

Fig. 3. Obtained weighted sum throughput and number of iterations for different approximation factor $\delta$

In Fig. 3, we plot the optimal weighted-sum throughput obtained by MAPEL, together with the needed number of iterations versus $\delta$. It is not surprising to see that the algorithm performance improves with a decreasing value $\delta$, which has been predicted by Theorem 3. On the other hand, the total number of iterations increases when $\delta$ decreases, and the change is drastic when $\delta$ is close to 0. Moreover, the algorithm performance is not sensitive to the value of $\delta$. For example, when $\delta = 0.1$, we achieve a weighted-sum throughput of 4.655bps/Hz that is only 0.025% away from the exact optimum. This illustrates that the performance bound obtained in Theorem 3 is quite loose, and actual performance can be much better. It is also clear that parameter $\delta$ provides a tuning knob for achieving desired trade-off between algorithm performance and computational complexity.

**Example 2** (Global optimal power allocation): MAPEL enables us to easily characterize the global optimal solution[13] of the WTM problem for an arbitrary wireless network. This is not possible before without exhaustive search. We consider a different 4-link network in Fig. 4 as a simple illustrating example. The length of each link is 4m, while the distances between $T_i$ to $R_j$ for $i \neq j$, denoted by $l_{ij}$, are proportional to $d$. The four links have different channel gains due to different fading states: $G_{11} = 1$, $G_{22} = 0.75$, $G_{33} = 0.50$, $G_{44} = 0.25$. The priority weight of each link is equal. Meanwhile, $G_{ij} = l_{ij}^{-4}$, $\boldsymbol{P}^{max}$=[0.7 0.8 0.9 1.0]mW, $n_i = 0.1\mu$W for all $i$. In Fig. 5, the optimal transmit power of each link is plotted against the topology parameter $d$. It can be seen that when the links are very close to each other, only the link with the largest channel gain (i.e., Link 1) is active with maximum transmit power $P_1^{max}$, while all the other links keep silent. When $d$ increases, a quantum jump in $p_2$ from 0 to $P_2^{max}$ is observed. As $d$ further increases, Link 3 starts to transmit, followed by Link 4. In this particular example, priority is always given to the link with a larger channel gain. Although the result may not be general, this toy example illustrates the possibility of using MAPEL as a tool to investigate the characteristics of global optimal solutions to power control problems.

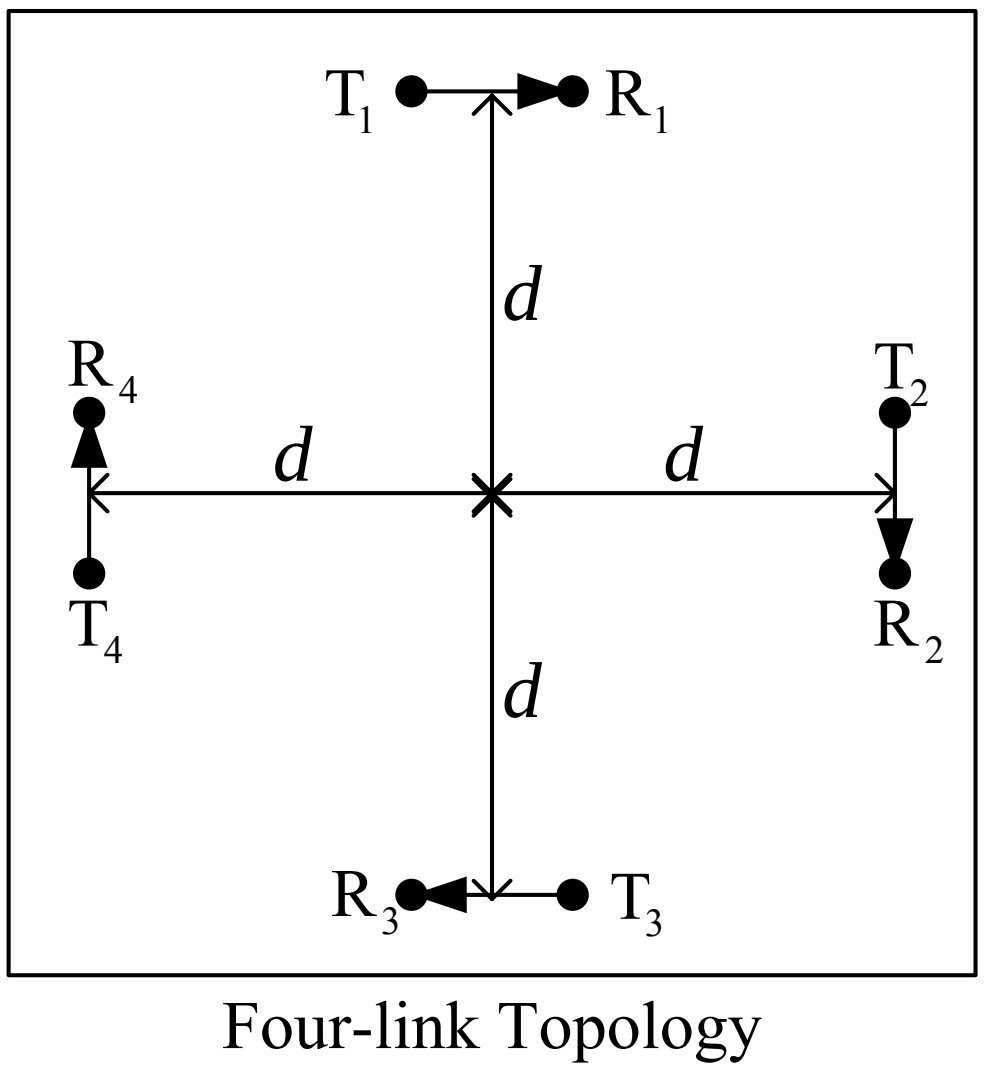

Fig. 4. The relationship between optimal transmit power and distance $d$

[13] MAPEL will only find one of the possible many global optimal solutions, depending on the choice of initial conditions.

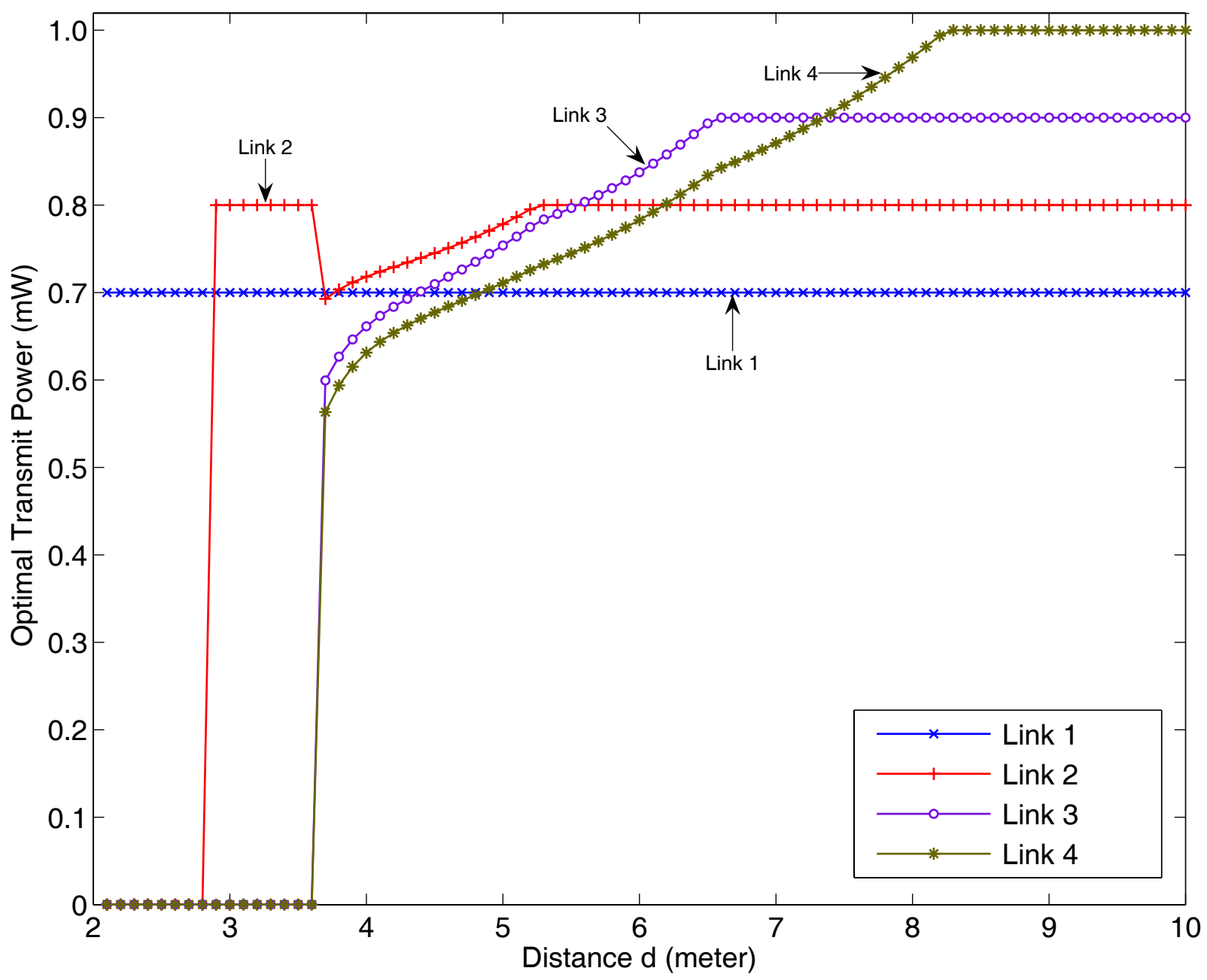

Fig. 5. A network topology with four links

## VI. Providing Benchmark for Existing Power Control Algorithms

A key application of MAPEL is to provide performance benchmark for other centralized or distributed algorithms that have been (or to be proposed) to solve WTM problem. With MAPEL, we are able to give quantitative measurements of these algorithms' performances (e.g., the chances of achieving global optimal solution and the gap of sub-optimality) under a wide range of network scenarios (e.g, different network densities and topologies).

### *A. Review of Existing Power Control Algorithms*

As we mentioned in Introduction, the current existing power control algorithms are essentially divided into two categories: centralized and distributed. Here we will review one "representative" algorithm from each category that represents the state-of-art in this area. Notice that the focus here is to show how MAPEL can be used to provide effective benchmark for the algorithms that tackle the same problem (i.e., Problem P1). Readers can choose your favorite algorithm to conduct the study.

*1) Centralized algorithm: Signomial Programming Condensation (SPC) Algorithm [12]:* SPC Algorithm is one of the best existing centralized algorithms for solving Problem P1. It utilizes the fact that Problem P1 can be rewritten as minimizing a ratio between two posynomials (i.e., a SP):

$$\begin{array}{ll} \text{minimize} & \prod_{i=1}^{M} \frac{g_i^{w_i}(\boldsymbol{p})}{f_i^{w_i}(\boldsymbol{p})} \\ \text{variables} & \boldsymbol{p} \in \mathcal{P} \end{array} \quad (15)$$

The key idea of SPC Algorithm is to improve the solution of Problem (15) through successive approximations until a KKT point is reached. During each step, the SP is approximated by a GP, which can be solved efficiently using a centralized interior point method.

*2) Asynchronous Distributed Pricing (ADP) Algorithm [16]:* ADP Algorithm is a distributed algorithm that can be used to solve Problem P1 without minimum data rate constraints. In ADP, each link announces a price that reflects its sensitivity to the received interference, and updates its own transmit power based on the prices announced by other links. The price and power values need to be updated iteratively and asynchronously until a convergent point is found. To implement the updates, each link only needs to acquire limited information from the network. We observe that ADP algorithm converges very fast in our numerical experiments, mainly because no stepsize is used in the updates. Its theoretical convergence to the global optimal point, however, is difficult to prove in general.

### *B. Performance Study of SPC Algorithm and ADP Algorithm*

In this subsection, we evaluate the performance of both algorithms through several examples by utilizing the benchmark provided by MAPEL.

**Example 3** (Probability of achieving global optimal solution): MAPEL always guarantees global optimality, while the SPC algorithm and the ADP algorithm fail to do so. Using the same 4-link network given in Example 1 (topology $\boldsymbol{G}_1$), we simulate three algorithms based on 500 different random initializations and show the results in Fig. 6 and Fig. 7, respectively. Then we change the topology to $\boldsymbol{G}_2$ with channel matrix illustrated in (16), and simulate both algorithms again in Fig. 8 and Fig. 9, respectively.

$$\boldsymbol{G}_2 = \begin{bmatrix} 0.1476 & 0.0105 & 0.0018 & 0.0402 \\ 0.0034 & 0.1784 & 0.0013 & 0.2472 \\ 0.0014 & 0.0017 & 0.3164 & 0.0046 \\ 0.0048 & 0.4526 & 0.0012 & 0.6290 \end{bmatrix}. \quad (16)$$

Other system parameters are the same as in Example 1. The figures show that MAPEL always converges to the global optimal solution, regardless of the initial power allocation. On the other hand, the SPC algorithm and the ADP algorithm are trapped in local optimal solutions from time to time. For example, Fig. 6 and Fig. 7 show that SPC and ADP algorithms obtain the global optimal solution 70.8% and 62.6% of the time, respectively. However, Fig. 8 and Fig. 9 show that in a different topology SPC and ADP algorithms obtain the global optimal solution 96% and 93.6% of the time, respectively.

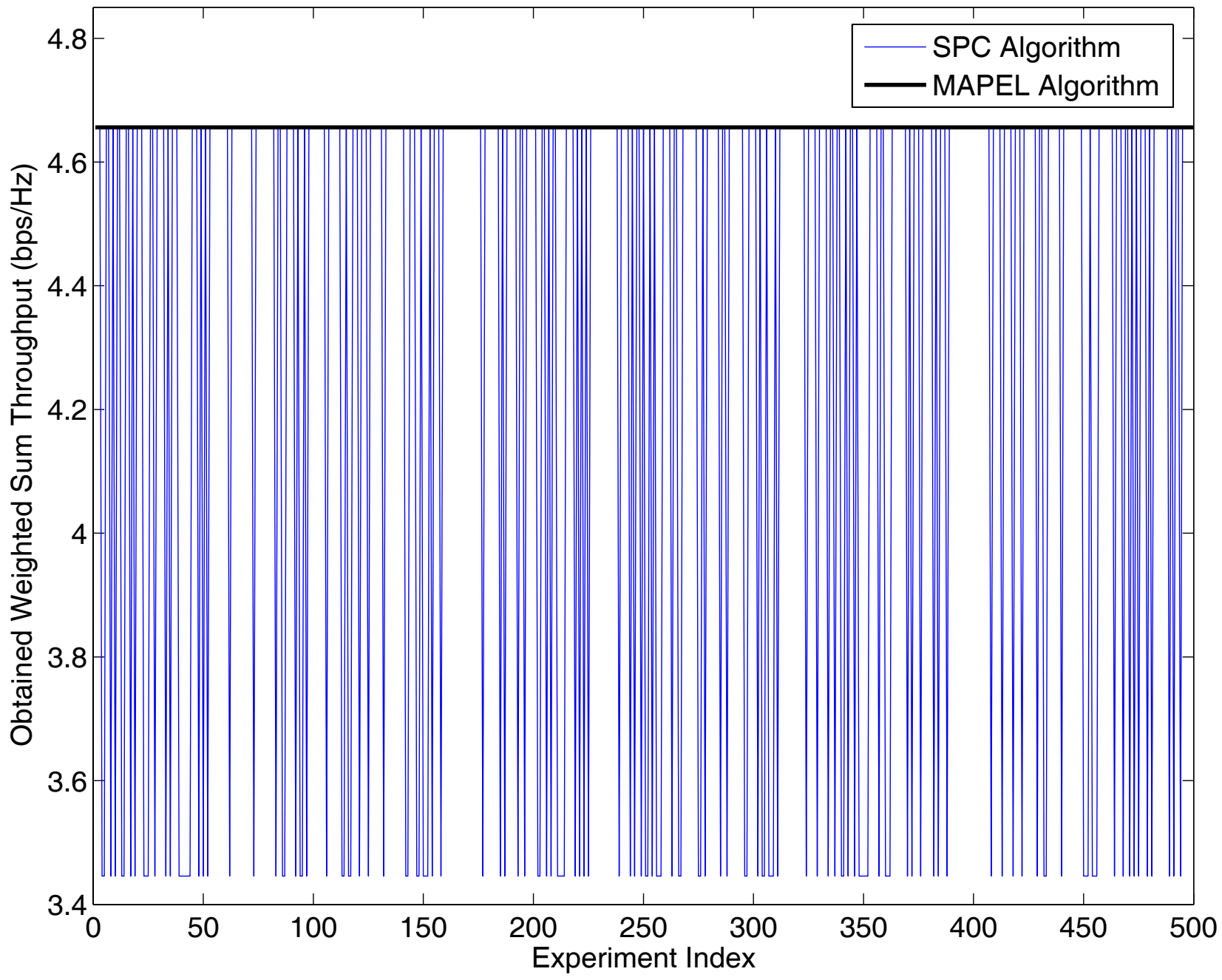

Fig. 6. Maximal weighted sum throughput achieved by MAPEL algorithm as well as SPC algorithm for 500 different initial feasible power allocations in $\boldsymbol{G}_1$ network

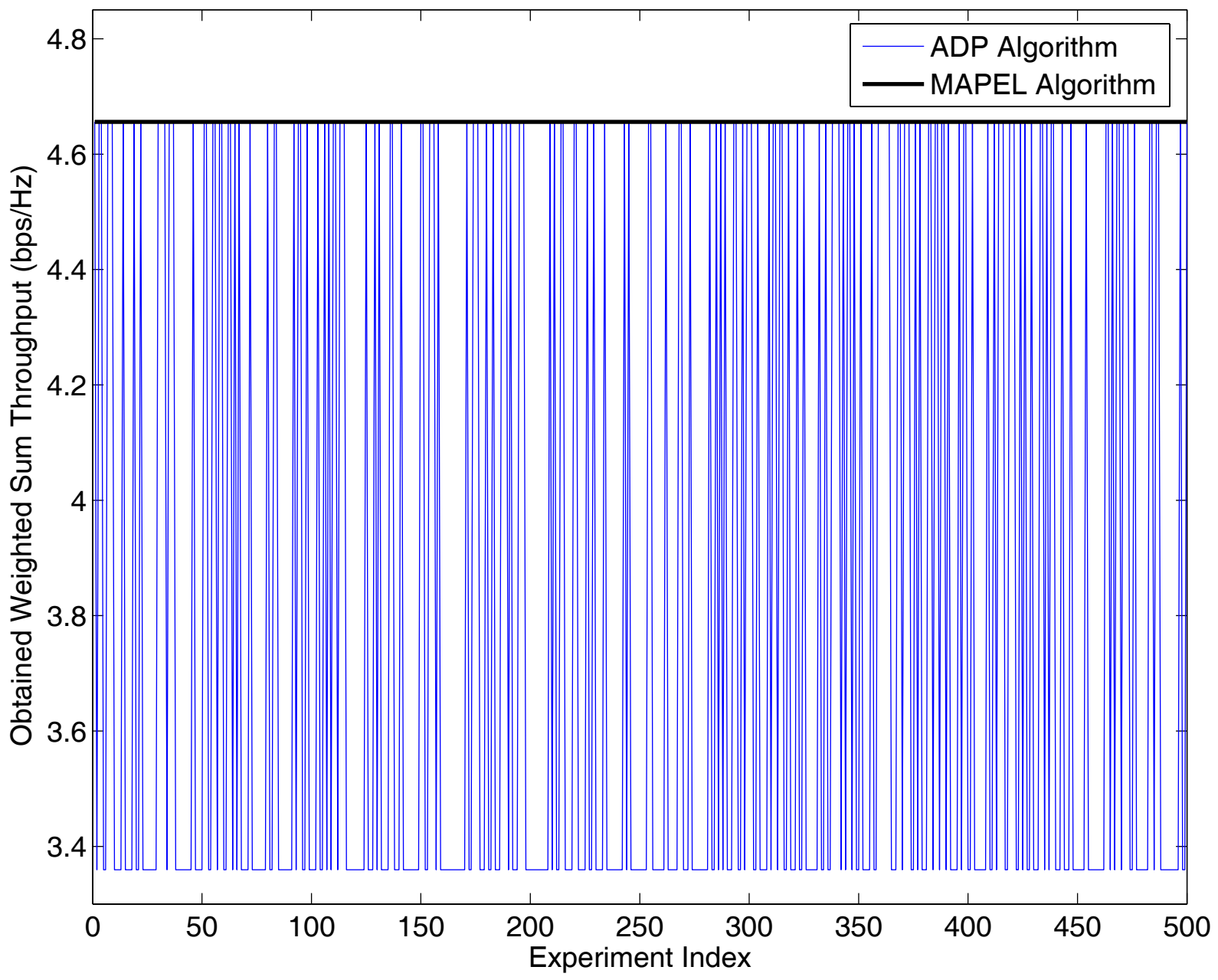

Fig. 7. Maximal weighted sum throughput achieved by MAPEL algorithm as well as ADP algorithm for 500 different initial feasible power allocations in $\boldsymbol{G}_1$ network

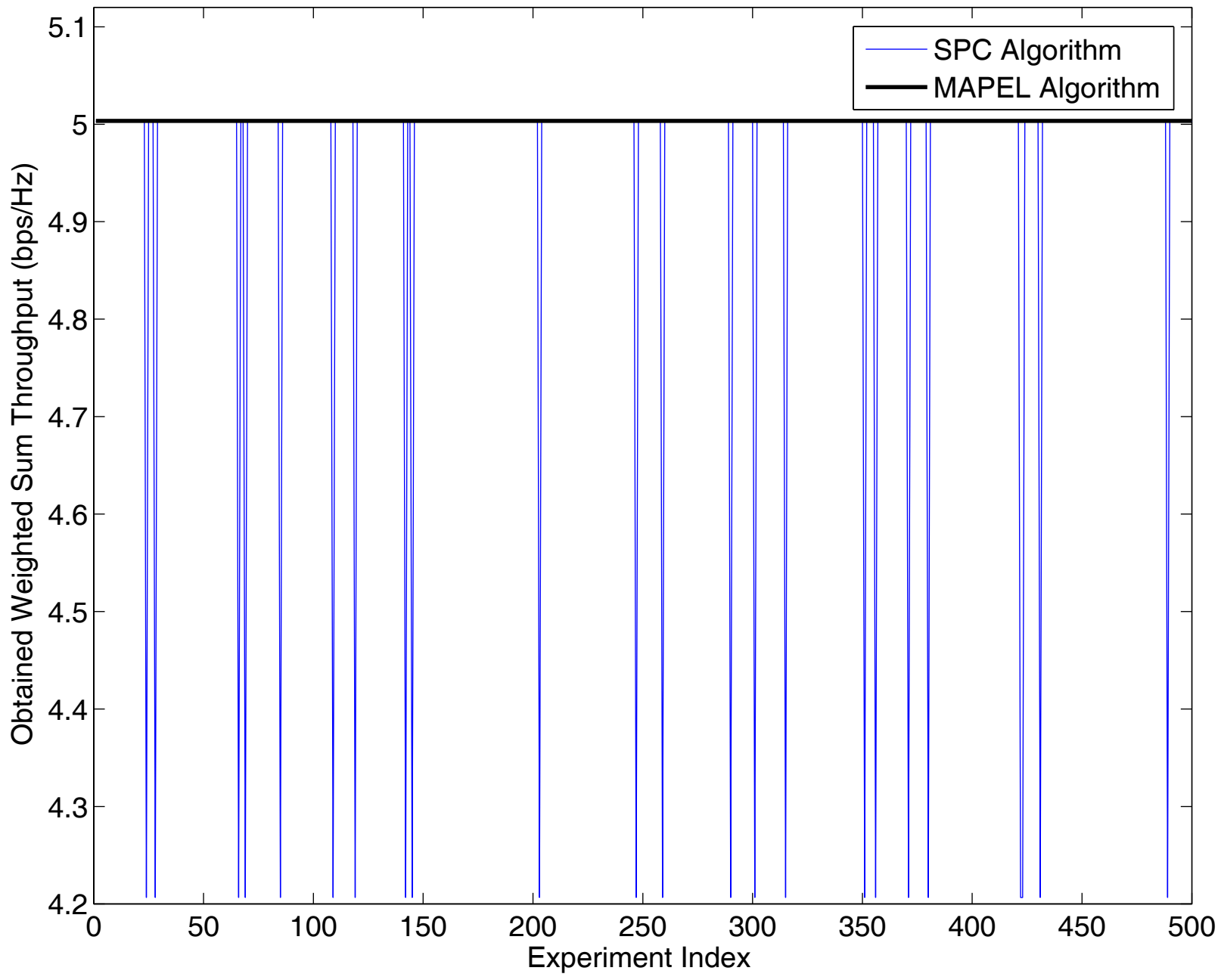

Fig. 8. Maximal weighted sum throughput achieved by MAPEL algorithm as well as SPC algorithm for 500 different initial feasible power allocations in $\boldsymbol{G}_2$ network

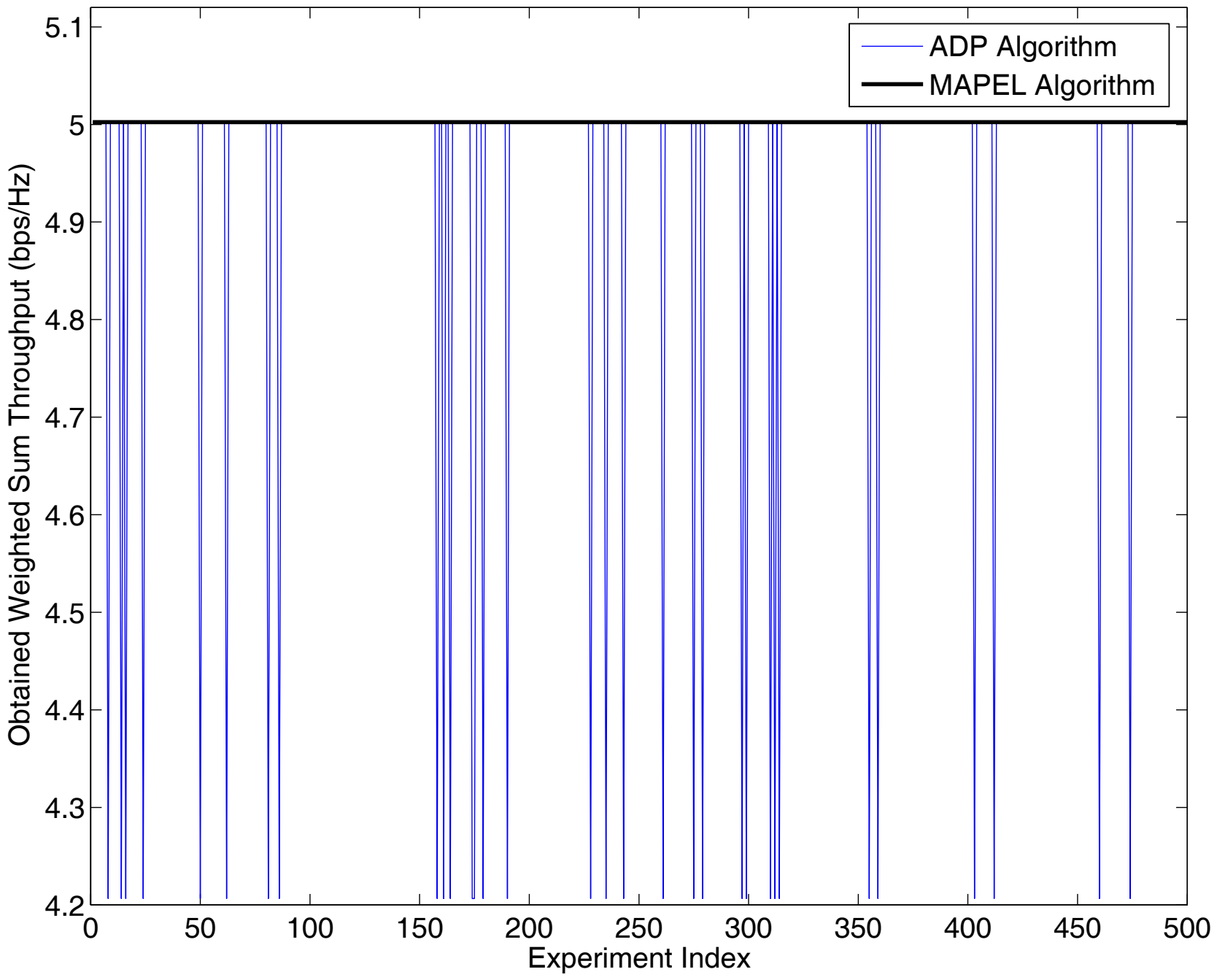

Fig. 9. Maximal weighted sum throughput achieved by MAPEL algorithm as well as ADP algorithm for 500 different initial feasible power allocations in $\boldsymbol{G}_2$ network

**Example 4** (Average algorithm performance without minimum data rate constraints): In Fig. 10, we compare the average performance of the SPC algorithm, the ADP algorithm, and the GP algorithm with MAPEL under different network densities. Compared with SPC and ARP, GP [12] approximates and solves the WTM problem based on high-SINR assumptions. For each fixed total number of links $n$, we place the links randomly in a 10m-by-10m area. The length of each link is uniformly distributed within [1m, 2m]. The priority weight of each link is equal. Meanwhile, we have $P_i^{max}$=1mW, $n_i = 0.1\mu$ W, and initial power allocation is fixed at $\boldsymbol{P}^{max}/2$. We vary the total number of links $n$ from 1 to 10. Each point is obtained by averaging over 500 different topologies of the same link density. On average, the performance loss of SPC compared with respect to the global optimality is about 2%, thus is quite small. Notice that the performance loss of each particular realization might be smaller (e.g., 0% when reaching the global optimality) or larger (when trapped in a local optimal). The average performance degradation of the ADP algorithm is about 10%, which implies that ADP is trapped in local optimum more often than SPC. Noticeably, the gap between SPC (or ADP) and the global optimum is not known before this work, as there does not exist previous algorithm that can guarantee the global optimal solution. This is in fact one of the key contributions of this paper. In addition, Fig. 10 shows that GP works reasonably well when the network density is low, where all (or most) links are active and some of them are indeed in the high SINR regime. However, the gap from the global optimum is much bigger when the network density becomes high, where many links need to be silent in order to avoid heavy interferences to their neighbors.

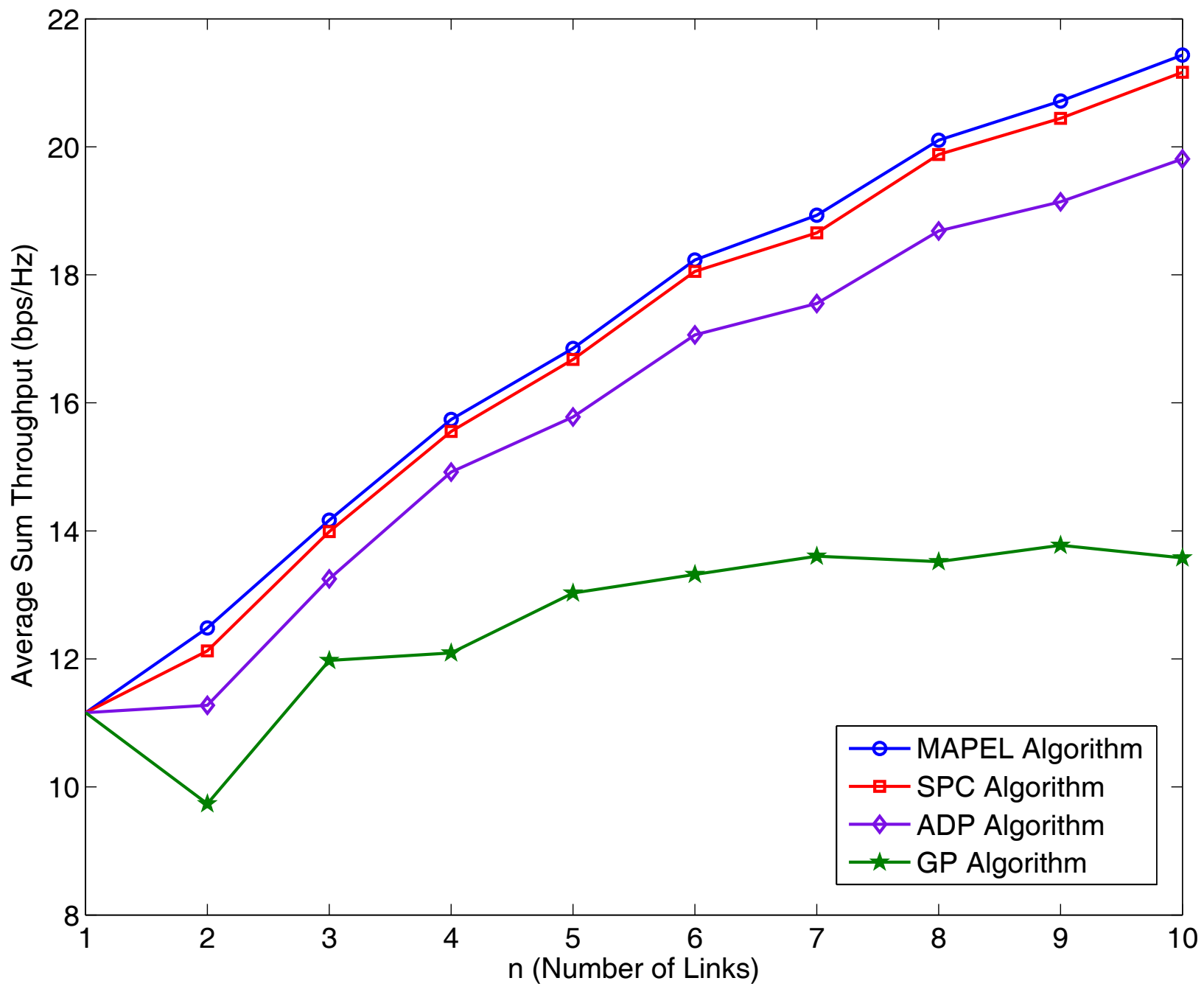

Fig. 10. Average sum throughput of different algorithms

Table I gives more detailed statistics about the performances of two algorithms. As shown in Table I, SPC achieves the global optimality with a probability that is always larger than 65% with the number of links up to 10. In contrast, the probability of ADP achieving the global optimality can be very low, e.g., only 0.6% in 10-link networks. It suggests that the initial power allocation of $\boldsymbol{P}^{max}/2$ is a good initial point for SPC, but may not for ADP. On the other hand, we find that SPC has a high-mean and low-variance average performance compared to the global optimality, which implies that SPC can achieve close-to-optimal performance with the initial power allocation of $\boldsymbol{P}^{max}/2$ for most topologies. However, ADP has a low-mean and high-variance average performance, which implies that ADP maintains a large degradation for some topologies.

**Example 5** (Average algorithm performance with minimum data rate constraints): We consider a series of 4-link networks with minimum data rate constraints on each link. The four links are randomly placed within a 10m x 10m area, and the length of each link is uniformly distributed within the interval [1m, 2m]. $\boldsymbol{P}^{max}$=[0.7 0.8 0.9 1.0]mW, $n_i = 0.1\mu$ W for all $i$. Meanwhile, the priority weight of each link is equal. In Fig. 11, the performance of MAPEL, GP, and SPC is plotted against the data rate constraint of each link. Each point for sum throughput on the curves is an average over 500 different topologies. We eliminate the topologies that are not feasible. Since ADP algorithm performs poorly in this case, we do not show its performance here.

It is not surprising to see that the sum throughputs of all algorithms drop as the data rate constraints become more stringent. One interesting observation is that the gap between GP and MAPEL becomes smaller when the data rate constraints are high. This is due to the fact that links are forced to operate in the high SINR regime when a high data rate is to be ensured. The high SINR assumption made by GP becomes more reasonable in this case.

TABLE I
OPTIMALITY OF SPC ALGORITHM AS WELL AS ADP ALGORITHM

| Number Of Links | SPC Algorithm | | | ADP Algorithm | | |
|---|---|---|---|---|---|---|
| | Probability of achieving global optimality | Average performance | Coefficient of variation | Probability of achieving global optimality | Average performance | Coefficient of variation |
| 2 | 69.8% | 96.9% | 7.22% | 50.6% | 89.6% | 17.8% |
| 4 | 80.4% | 98.7% | 3.91% | 25.0% | 94.3% | 8.79% |
| 6 | 77.2% | 98.9% | 3.13% | 6.0% | 93.4% | 7.89% |
| 8 | 69.4% | 98.8% | 2.58% | 1.4% | 92.7% | 7.42% |
| 10 | 65.6% | 98.7% | 2.81% | 0.6% | 92.1% | 8.18% |

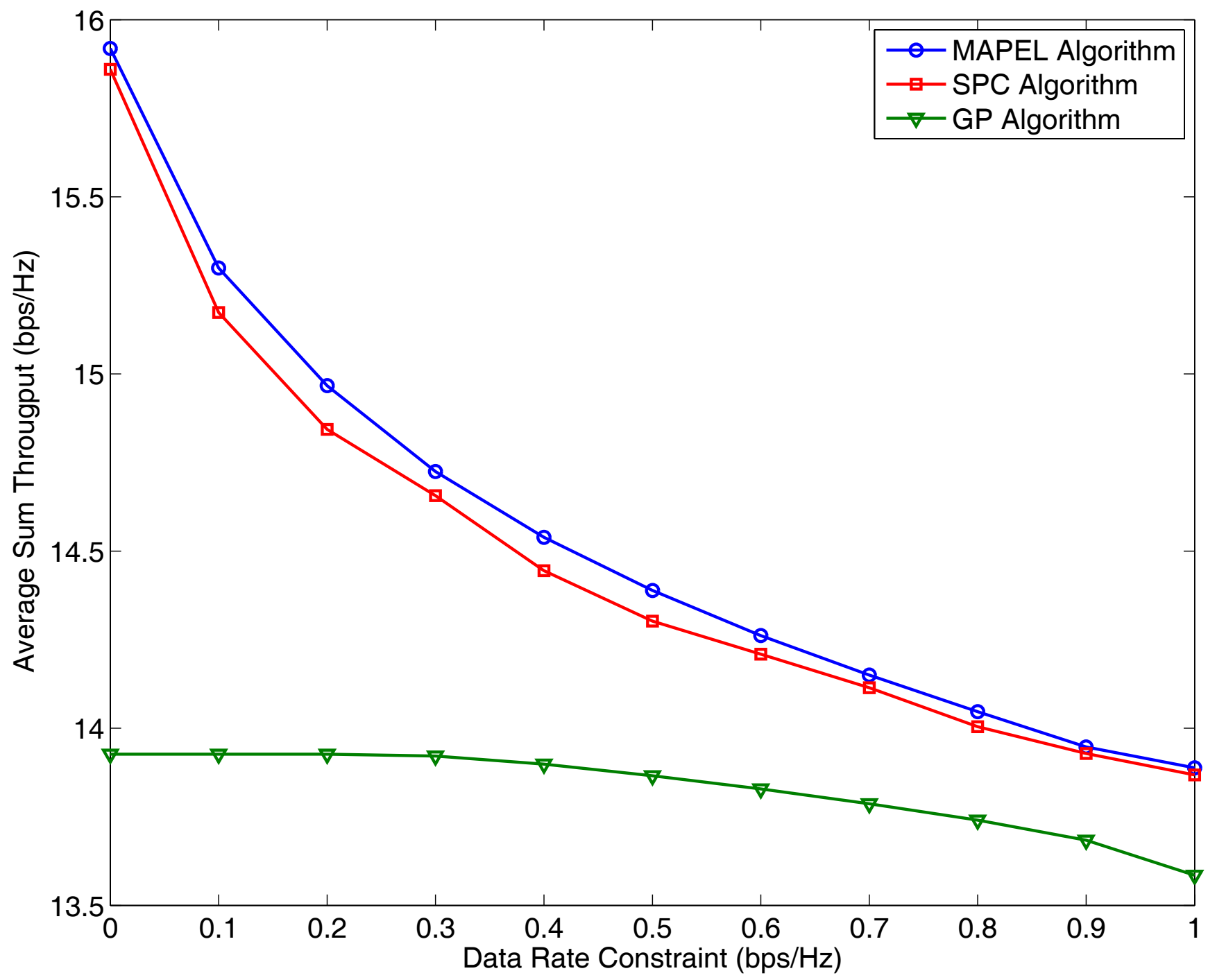

Fig. 11. Average sum throughput of different algorithms versus the data rate constraint in 4-link networks

## VII. Conclusions and Discussions

In this paper, we proposed the MAPEL algorithm that solves the open problem of weighted throughput maximization in general interference-limited wireless networks. The MAPEL algorithm is guaranteed to globally converge to an optimal solution despite the nonconvexity of the problem. The key idea behind the algorithm is to reformulate the WTM problem into an MLFP and then construct a sequence of shrinking polyblocks that eventually closely approximate the upper boundary of the feasible region around the global optimum. We have also established the tradeoff relationship between performance and convergence time of the MAPEL algorithm.

Although a centralized algorithm, MAPEL provides an important benchmark for performance evaluation of existing and newly proposed power control heuristics in this area. For example, by comparing with MAPEL through extensive simulations, we have gained deeper understanding of two state-of-the-art centralized and distributed power control algorithms: SPC algorithm and ADP algorithm. Simulations show that both algorithms achieve close-to-optimal average performance in the general SINR regime.

This paper helps to pave the way for further study of power control problems with various objectives and constraints. An interesting future research direction is to study power control that maximizes general utility functions, including both concave and non-concave functions. Optimal power control in time-varying channels is another challenging topic for future research.

The MAPEL algorithm presented in this paper is not the only way to efficiently obtain the global optimal solution. Variants of the algorithm can be developed to expedite the convergence and reduce the computational complexity. For example, it can be proved that the projection of a vertex of $\mathcal{S}$ on $\mathcal{G}$ must contain at least one element equal to $P_i^{max}$. Such characteristics could be used to design a faster projection algorithm to replace Algorithm 1. Another possibility is to exploit the shape of the feasible region $\mathcal{G}$.

## REFERENCES

[1] M. Chiang, P. Hande, T. Lan and C. W. Tee, "Power Control Wireless Cellular Networks," to appear in *Foundations and Trends in Networking*, 2008

[2] J. Zander, "Performance of optimum transmitter power control in cellular radio systems," in *IEEE Trans. Veh. Technol.*, vol. 41, no. 1, pp. 57-62, February 1992.

[3] G. J. Foschini and Z. Miljanic, "A simple distributed autonomous power control algorithm and its convergence," in *IEEE Trans. Veh. Technol.*, vol. 42, no. 6, pp. 641-646, November 1993.

[4] S. A. Grandhi and J. Zander, "Constrained Power Control in Cellular Radio Systems," *Proc. IEEE VTC*, Stockholm, Sweden, June 1994.

[5] S. A. Grandhi, R. Vijayan, D. J. Goodman, and J. Zander, "Centralized Power Control in Cellular Radio Systems," *IEEE Trans. Veh. Technol.*, vol. 42, no. 6, pp. 466-468, November 1993.

[6] R. D. Yates, "A Framework for Uplink Power Control in Cellular Radio Systems," *IEEE J. Sel. Areas Commun.*, vol. 13, no. 7, pp. 1341-1347, September 1995.

[7] N. Bambos, S. C. Chen, and G. J. Pottie, "Channel access algorithms with active link protection for wireless communication networks with power control," in *IEEE/ACM Trans. Netw.*, vol. 8, no. 5, pp. 583-597, October 2000.

[8] C. W. Sung, "Log-Convexity Property of the Feasible SIR Region in Power-Controlled Cellular Systems," *IEEE Commun.Letters*, vol. 6, no. 6, pp 248-249, June 2002.

[9] H. Boche and S. Stanczak,"Convexity of some feasible QoS regions and asymptotic behavior of the minimum total power in CDMA systems, *IEEE Trans. Commun.*, vol. 52, no. 12, pp. 2190 - 2197, December 2004.

[10] M. Schubert and H. Boche, "Solution of the Multi-user Downlink Beamforming Problem with Individual Sir Constraints," *IEEE Trans. Veh. Technol.*, vol. 53, no. 1, pp. 18-28, January 2004.

[11] D. Julian, M. Chiang, D. O. Neill, and S. Boyd, "Qos and fairness constrained convex optimization of resource allocation for wireless cellular and ad hoc networks," *Proc. IEEE INFOCOM*, vol. 2, pp. 477-486, June 2002.

[12] M. Chiang, C. W. Tan, D. Palomar, D. O'Neill, and D. Julian, "Power control by geometric programming," *IEEE Trans. Commun.*, vol. l, no. 7, pp. 2640-2651, July 2007.

[13] M. Chiang and J. Bell, "Balancing supply and demand of wireless bandwidth: Joint rate allocation and power control," *Proc. IEEE INFOCOM*, Hong Kong, China, March 2004.

[14] D. ONeill, D. Julian, and S. Boyd, “Adaptive management of network resources,” *Proc. IEEE VTC*, Orlando, FL, October 2003.

[15] M. Xiao, N. B. Shroff, and E. Chong, “A utility-based power-control scheme in wireless cellular systems,” *IEEE/ACM Trans. Networking*, vol. 11, no. 2, pp. 210 - 221, April 2003.

[16] J. Huang, R. A. Berry, and M. L. Honig, “Distributed Interference Compensation for Wireless Networks,” *IEEE J. Sel. Areas Commun.*, vol. 24, no. 5, pp. 1074 - 1084, May 2006.

[17] P. Hande, S. Rangan, and M. Chiang, “Distributed uplink power control for optimal sir assignment in cellular data networks,” *Proc. IEEE INFOCOM*, April 2006.

[18] M. Chiang, “Balancing transport and physical layers in wireless multihop networks: Jointly optimal congestion control and power control,” *IEEE J. Sel. Areas Commun.*, vol. 23, no. 1, pp. 104-116, January 2005.

[19] X. Lin, Ness B. Shroff and R. Srikant, “A Tutorial on Cross-Layer Optimization in Wireless Networks,” *IEEE J. Sel. Areas Commun.*, vol. 24, no. 8, pp. 1452-1463, August 2006.

[20] N. T. H. Phuong and H. Tuy, “A Unified Monotonic Approach to Generalized Linear Fractional Programming,” Journal of Global Optimization, Kluwer Academic Publishers, pp. 229-259, 2003.

[21] J. B. G. Frenk and S. Schaible, “Fractional Programming,” Handbook of Generalized Convexity and Generalized Monotonicity, pp. 335-386, 2006.

[22] S. Boyd and L. Vandenberghe, “*Convex Optimization*,” Cambridge, U.K.: Cambridge Univ. Press, 2004.